\documentclass[prd, preprint, showpacs,amsmath,amssymb]{revtex4-1}
\usepackage{latexsym}
\usepackage{graphics}
\usepackage{epsfig}
\usepackage{dcolumn}
\def\br{\begin{eqnarray}}
\def\er{\end{eqnarray}}
\def\be{\begin{equation}}
\def\ee{\end{equation}}

\def\({\left(}
\def\){\right)}
\def\<{\left\langle}
\def\>{\right\rangle}

\begin{document}
%
%
\title{ The Gauge Bosons Masses in a $SU(2)_{TC}\otimes SU(3)_{{}_{L}}\otimes U(1)_{{}_{X}}$ Extension of the Standard Model}
\author{A. Doff$^{a}$} 
\affiliation{$^a${\small Universidade  Tecnol\'ogica  Federal do Paran\'a - UTFPR - COMAT, Via do Conhecimento Km 01, 85503-390, Pato  Branco, PR, Brazil}}

\date{\today}
\begin{abstract} 
 The gauge symmetry breaking in 3-3-1 models can be implemented dynamically because at the scale of a few TeVs  the $U(1)_X$
coupling constant becomes strong.  The exotic quark $T$  introduced in the model will form a condensate  breaking  $SU(3)_{{}_{L}}\otimes U(1)_X$ to  electroweak symmetry. In this brief report  we explore the full  realization of the  dynamical symmetry breaking of an 3-3-1 model  to $U(1)_{em}$  considering a model based on  $SU(2)_{TC}\otimes SU(3)_{{}_{L}}\otimes U(1)_{{}_{X}}$. We compute the mass generated for the charged  and neutral gauge bosons of the model that result from the symmetry breaking, and  verify  the equivalence between a  3-3-1 model  with a  scalar content formed by the set of the fundamental scalar bosons $\chi, \rho$ and $\eta$   with a 3-3-1 model where the   dynamical symmetry breaking  is implanted by the system formed by the set of composite bosons $\Phi_{{}_{T}},  \Phi_{{}_{TC(1)}}$ and  $\Phi_{{}_{TC(2)}}$. In this model the minimal composite scalar content  is  fixed by the  condition  of the cancellation of  triangular  anomaly in TC sector.
\end{abstract}
\maketitle

  The Standard Model of elementary particles is in excellent agreement with the experimental data and has explained
many features of particle physics throughout the years. Despite its success, there are some points in the
model that could be better explained with the introduction of new fields and symmetries, such as the flavor problem or
the enormous range of masses between the lightest and heaviest fermions and other peculiarities. One of the possibilities in this direction is to assume an extension of the standard model based on $G_{3n1} \equiv  SU(3)_{{}_{C}}\otimes SU(n)_{{}_{L}}\otimes U(1)_{{}_{X}}$\cite{felice1, frampton, tonasse, felice2}, where $n=3,4$. This class of the models predicts interesting new physics at TeV scale\cite{trecentes} and addresses some fundamental questions that cannot be explained in the framework of the Standard Model. As a brief example we can mention the flavor problem\cite{dp1} and the question of the electric charge quantization\cite{dp2}. 

 One interesting feature of some versions of these models  is the following relationship among the coupling
constants $g$ and $g'$ associated to the gauge group $SU(3)_{{}_{L}}\otimes U(1)_{{}_{X}}$ 
\begin{equation}
\frac{\alpha'}{\alpha} = \frac{\sin^2\theta_{{}_{W}}(\mu)}{1 - 4\sin^2\theta_{{}_{W}}(\mu)}
\label{eq1}
\end{equation}
\noindent where $\alpha = g^2/4\pi$, $\alpha' = g'^2/4\pi$   and  $\theta{{}_{W}}$ is the electroweak mixing angle. As argued in  Refs.\cite{Das, doff}, the gauge symmetry breaking of $SU(3)_{{}_{L}}\otimes U(1)_{{}_{X}}$ in 3-3-1 models can be implemented dynamically because at the scale of a few TeVs, $\mu_X$, the $U(1)_X$ coupling constant $(g')$  becomes strong as we approach the peak existent in Eq.(\ref{eq1}). The exotic quark $J_3$ introduced in these models, in our notation $J_3 \equiv T$ ,  will form a condensate  breaking  $SU(3)_{{}_{L}}\otimes U(1)_{{}_{X}}$  to electroweak gauge symmetry  without requiring the introduction of fundamental scalars.  In Ref.\cite{doff} we investigated  this possibility  and  show that just  version\cite{tonasse} of this class of models leads  to a deeper minimum of the effective potential.

The mechanism that breaks the electroweak  symmetry $SU(2)_{{}_{L}}\otimes U(1)_{{}_{Y}}$  down to the gauge symmetry of electromagnetism
$U(1)_{em}$  is still the only obscure part of the standard model  and  the  understanding of the gauge electroweak symmetry breaking mechanism is one of the most important problems in particle physics at present. One of the explanations of this mechanism  is based on the introduction  of a new strong interaction  usually named technicolor (TC), where in these theories the Higgs boson is a composite of the so called technifermions. The beautiful characteristics of technicolor (TC) as well as its problems are clearly described in  Refs.\cite{lane, simmons}. 

\par In this work we will extend  the results obtained in \cite{doff}, in particular,  we intend to explore the full realization of the  dynamical symmetry breaking of  model\cite{tonasse}  to $U(1)_{em}$  considering   $SU(2)_{TC}\otimes SU(3)_{{}_{L}}\otimes U(1)_{{}_{X}}$  group, where $SU(2)_{TC}$ is the minimal TC gauge group that will be responsible for the electroweak symmetry breaking. 

   We begin determining the mass generated for the charged  gauge bosons of the model that results from the symmetry breaking assuming the charged current interactions  associated to the technifermions and to  the exotic quark T, that  will be  responsible for the mass generation of the 
 heavy gauge bosons  $V^{\pm}$ and $U^{\pm\pm}$. In the sequence we  obtain the  neutral current interactions and the  mass matrix generated for the neutral gauge bosons. 
    
  The fermionic content of the model  has the same quark sector  of Ref.\cite{felice1} 
\br 
&& Q_{3L} = \left(\begin{array}{c} t \\ b \\ T  \end{array}\right)_{L}\,\,\sim\,\,({\bf 1}, {\bf 3}, 2/3) \nonumber \\ \nonumber \\
&&t_{R}\,\sim\,({\bf 1}, {\bf 1}, 2/3)\,,\,b_{R}\,\sim\,({\bf 1},{\bf 1},-1/3)\nonumber  \\  \nonumber \\
&&T_{R}\,\sim\, ({\bf 1},{\bf 1},  5/3) \nonumber 
\er
\br
&&Q_{\alpha L} = \left(\begin{array}{c} D \\ u \\ d  \end{array}\right)_{\alpha L}\,\,\sim\,\,({\bf 1}, {\bf 3^*}, -1/3) 
\nonumber \\  \nonumber \\
&&u_{\alpha R}\, \sim\, ({\bf 1},{\bf 1},  2/3 )\,,\,d_{\alpha R}\,\sim\,({\bf 1}, {\bf 1}, -1/3)\nonumber  \\   \nonumber \\
&&D_{\alpha R}\,\sim\, ({\bf 1}, {\bf 1},  -4/3 )
\er
\noindent  where $\alpha = 1,2 $ is the family index and  we represent  the third quark  family  by $Q_{3L}$. In these expressions $({\bf 1},{\bf 3} ,  X)$, $({\bf 1}, {\bf 3^*},  X)$ or $({\bf 1}, {\bf 1},  X)$ denote the  transformation properties  under  $SU(2)_{TC}\otimes SU(3)_{{}_{L}}\otimes U(1)_{{}_{X}}$ and $X$ is the  corresponding $U(1)_{X}$ charge.  The leptonic sector includes beside the  conventional  charged leptons and their respective neutrinos, charged heavy leptons $E_a$\cite{tonasse}.  
\br
&& l_{aL} = \left(\begin{array}{c} \nu_{a} \\ l_a \\ E^c_a\end{array}\right)_{L}\,\sim\,({\bf 1},{\bf 3},  0)
\er
\noindent where $a=1,2,3$ is the family index and $l_{aL}$ transforms as triplets   under $SU(3)_L$. Moreover,  we have to add the corresponding right-handed components, $l_{aR} \sim ({\bf 1},{\bf 1}, -1)$ and $E^c_{aR} \sim ({\bf 1},{\bf 1},+1)$.  
\par In addition, we  included the minimal Technicolor  sector, represented by
\br 
&&\Psi_{1L} = \left(\begin{array}{c} U_1\\ D_1\\  U'\end{array}\right)_{L}\,\,\sim\,\,({\bf 2}, {\bf 3}, 1/2) \nonumber  \\ 
\nonumber \\ 
&&U_{1 R}\, \sim\, ({\bf 2}, {\bf 1}, 1/2)\,,\,D_{1 R}\, \sim \,({\bf 2}, {\bf 1},-1/2)\nonumber \\
&&U'_{R}\,\sim\, ({\bf 2},{\bf 1}, 3/2)\nonumber  ,
\er
\br 
&&\Psi_{2L} = \left(\begin{array}{c} D'\\ U_2\\  D_2 \end{array}\right)_{L}\,\,\sim\,\,({\bf 2}, {\bf 3^*}, -1/2) \nonumber \\ 
\nonumber \\ 
&&U_{2 R}\, \sim\, ({\bf 2}, {\bf 1}, 1/2)\,,\,D_{2 R}\,\sim\,({\bf 2}, {\bf 1},-1/2)\nonumber \\ 
&&D'_{R}\,\sim\, ({\bf 2},{\bf 1}, -3/2).
\er
\noindent where $1$ and $2$ label the  first and second techniquark families, $U'$ and $D'$ can be considered as exotic techniquarks making an  analogy with quarks $T$ and $D$ that appear in the   fermionic content  of the model . The model is anomaly free if we have equal numbers of triplets and antitriplets, counting the color of $SU(3)_c$. Therefore, in order to make the model anomaly free two of the three quark
generations transform as ${\bf 3^*}$,  the third quark  family and the three leptons generations  transform as ${\bf 3}$.  It is easy to check that all gauge anomalies cancel out in this model,  in the TC sector the triangular anomaly cancels between the two generations of technifermions.  In the present version of the model  we assumed that technifermions are singlets of $SU(3)_c$.

The charged current interactions  with the  quark $T$   are described by 

\br
{\cal L}^{cc}_{Q_{3L}}  = \frac{g}{\sqrt{2}}\left(\bar{T}_L \gamma^{\mu} t_L V^{+}_{\mu} + \bar{T}_L \gamma^{\mu} b_L U^{++}_{\mu} + h.c.\right)
\er
\par For the   TC sector,  the charged current interactions to the first technifermion generation can  be written as  
\br 
{\cal L}^{cc}_{TC(1)}   = \frac{g}{\sqrt{2}}&&\left( \bar{U}_{1L} \gamma^{\mu} D_{1L} W^{+}_{\mu} + \bar{U'}_L \gamma^{\mu} U_{1L} V^{+}_{\mu}  + \right. \nonumber \\ &&\left. \bar{U'}_L \gamma^{\mu} D_{1L} U^{++}_{\mu} + h.c.\right),
\er
\noindent  while for the  second technifermion  generation we obtain  
\br
{\cal L}^{cc}_{TC(2)}  = \frac{g}{\sqrt{2}}&&\left( \bar{U}_{2L} \gamma^{\mu} D_{2L} W^{+}_{\mu} + \bar{D}_{2L} \gamma^{\mu} D'_{L} V^{+}_{\mu} +  \right. \nonumber \\ &&\left.  \bar{U}_{2L} \gamma^{\mu} D'_{L} U^{++}_{\mu} + h.c.\right). 
\er 
\noindent From these, we can extract the couplings of charged gauge bosons with the axial currents $J^{\mu}_{5(T)} = \frac{1}{2}\bar{T}\gamma^{\mu}\gamma_5\Psi_i$, with  $\Psi_i = t, b$,  $J^{\mu}_{5(1)} = \frac{1}{2}\bar{U_1}\gamma^{\mu}\gamma_5 D_1$,  $J^{\mu}_{5(1')} = \frac{1}{2}\bar{U'}\gamma^{\mu}\gamma_5 \Psi_j$ where $\Psi_j = U_1, D_1$, and  $J^{\mu}_{5(2)} = \frac{1}{2}\bar{U_2}\gamma^{\mu}\gamma_5 D_2$ , $J^{\mu}_{5(2')} = \frac{1}{2}\bar{\Psi_k}\gamma^{\mu}\gamma_5 D'$  to  $\Psi_k =  D_2, U_2$.  After  considering the decay constant  relations for  the axial currents  
\br 
&&\langle 0|J^{\mu}_{5(T)}|\Pi \rangle \sim i\frac{F_{\Pi}}{\sqrt{2}}p^{\mu} \nonumber \\
&& \langle 0|J^{\mu}_{5(1, 1')}|\pi_1 \rangle \sim i\frac{f_{\pi_{1}}}{\sqrt{2}}p^{\mu}\,,\,\langle 0|J^{\mu}_{5(2,2')}|\pi_2 \rangle \sim i\frac{f_{\pi_{2}}}{\sqrt{2}}p^{\mu} 
\er
\noindent we can write the interaction terms of the charged bosons $V^{\pm}$, $U^{\pm\pm}$ and $W^{\pm}$   with $U(1)_X$  and TC pions $(\Pi, \pi_1 ,\pi_2)$ as
\br
&&\hspace{-0.5cm}{\cal L}_{\Pi-V} = -\frac{ig}{2}F_{\Pi}p^{\mu}V^{\pm}_{\mu}\,\,,\,\, {\cal L}_{\pi-V} = -\frac{ig}{2}\left[f_{\pi_{1}}p^{\mu} + f_{\pi_{2}}p^{\mu}\right]V^{\pm}_{\mu}\nonumber \\
&&\hspace{-0.5cm}{\cal L}_{\Pi-U} = -\frac{ig}{2}F_{\Pi}p^{\mu}U^{\pm\pm}_{\mu}\!\!\!\!\,\,,\,\,{\cal L}_{\pi-U} = -\frac{ig}{2}\left[f_{\pi_{1}}p^{\mu} +  f_{\pi_{2}}p^{\mu}\right]U^{\pm\pm}_{\mu}\nonumber \\
&&\hspace{-0.5cm}{\cal L}_{\pi-W} = -\frac{ig}{2}\left[ f_{\pi_{1}}p^{\mu} + f_{\pi_{2}}p^{\mu}\right]W^{\pm}_{\mu}.
\label{chaco} 
\er 
\noindent In the equations  above the technipion decay constants, $( f_{\pi_{1}} = f_{\pi_{1}^{\pm}})$ and $(f_{\pi_{2}} = f_{\pi_{2}^{\pm}})$, are  related to the vacuum expectation value(VEV) of the Standard Model through
\be
 \left( f^2_{\pi_{1}} + f^2_{\pi_{2}} \right)  = v^2  = \frac{4M_W^2}{g^2} 
\ee
\begin{figure}[t]
\begin{center}
\epsfig{file=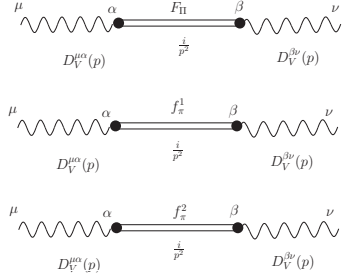,width=0.3\textwidth}
\caption{Contributions to the vacuum polarization $\Pi_{\alpha\beta}(p^2)$  of the charged gauge boson  $V^{\pm}$.}
\end{center}
\end{figure}
\noindent and we will consider that $F_{\Pi}\!\sim \!\mu_{X}\!\sim\!O(TeV)$. 

Technicolor models  with fermions in the fundamental representation  are subjected to a  strong experimental constraint that comes from
the limits on the $S$ parameter. In our case,  the contribution due to the  TC sector should still lead to a value to the S parameter compatible with the experimental data.  At low energies, i.e.  at the scale associated with electroweak symmetry breaking, we should only consider the contribution of four techniquarks because (U 'and D') are singlets of $SU(2)_{L}$ and do not contribute directly to the mass of (W and Z) bosons.

\par  In Fig.I   we  show the couplings in $O(g^2)$  between the charged pions, $\Pi^{\pm}$ and $\pi^{\pm}_{1,2}$, with the charged boson 
$V^{\pm}$. From this figure we can write the correction  to the  $V^{\pm}$ propagator as 
\br 
iD'_V(p^2)^{\mu\nu}\!\! = \! iD_V(p^2)^{\mu\nu}\! + \!i\frac{g^2}{2}D_V(p^2)^{\mu\alpha}\!\!\left[i\Pi^{V}_{\alpha\beta}(p^2) \right]iD_V(p^2)^{\beta\nu}\nonumber 
\er 
\noindent where $D_V(p^2)$ is the tree level  propagator in the Landau gauge  and $\Pi^{V}_{\alpha\beta}(p^2)$ is obtained from the pions couplings. Then, after considering  $\Pi^{V}_{\alpha\beta}(p^2) = \left(p^2g_{\alpha\beta} - p_\alpha p_\beta\right)\Pi^{V}(p^2)$ , $M^2_{V} = g^2 p^2\Pi^{V}(p^2)/2$,  the contributions for the  polarization tensor depicted in  Fig. I,  and the first equation listed in  (\ref{chaco}),  we obtain 
\be 
M^2_{V}  = \frac{g^2}{4}\left(F^2_{\Pi} + f^2_{\pi_{1}} + f^2_{\pi_{2}}  \right).
\ee 
\par  The mass generated for the  $U$ and $W$ bosons  can be obtained in the same way, these results are presented below  
\be
M^2_{U}  = \frac{g^2}{4}\left(F^2_{\Pi} + f^2_{\pi_{1}} + f^2_{\pi_{2}} \right)\,\,,\,\, M^2_{W} =  \frac{g^2}{4}\left( f^2_{\pi_{1}} + f^2_{\pi_{2}}\right). 
\ee
\par The mass generated for neutral bosons $Z_0$ and $Z'_0$ can be determined in a similar way, below we show the 
 couplings of exotic quark  $T$ with $W_8$  and $B$  
\br 
{\cal L}_{T-B-W^8} = &&+ \frac{g}{\sqrt{3}}\bar{T}_{L}\gamma_{\mu}T_{L}W^{\mu}_{8}   - \frac{2g'}{3}\bar{T}_{L}\gamma_{\mu}T_{L}B^{\mu}\nonumber \\ 
&&  - \frac{5g'}{3}\bar{T}_{R}\gamma_{\mu}T_{R}B^{\mu}
\er 
\noindent where $(B, W_3, W_8)$ are  symmetry eigenstates,  $B$ is the $U(1)_X$ boson and the eigenstates $(W_3, W_8)$  are associated to the neutral generators of the $SU(3)_L$.  For the first technifermion generation $(1^a ger)$, the  respective  couplings are listed below  
\br 
 {\cal L}^{1^a\, ger}_{W^3,W^8} = -\frac{g}{2}&&\left[ \bar{U}_{1L}\gamma_{\mu}U_{1L}\left(W^{\mu}_{3}  + \frac{1}{\sqrt{3}}W^{\mu}_{8} \right)  +  \right. \nonumber \\ &&\left. \bar{D}_{1L}\gamma_{\mu}D_{1L}\left(-W^{\mu}_{3}  + \frac{1}{\sqrt{3}}W^{\mu}_{8} \right) +  \right. \nonumber \\ 
&&\left.  -\frac{2}{\sqrt{3}}\bar{U'}_{L}\gamma_{\mu}U'_{L}W^{\mu}_{8} \right] 
\er 
\br 
{\cal L}^{1^a\, ger}_{B} = -\frac{g'}{2}&&\left[\bar{U}_{1L_{{}_{}}}\gamma_{\mu}U_{1L} + \bar{D}_{1L}\gamma_{\mu}D_{1L} + \right. \nonumber \\ &&\left. \bar{U}_{1R}\gamma_{\mu}U_{1R}  - \bar{D}_{1R}\gamma_{\mu}D_{1R}  +  \right. \nonumber \\
&&\left.  3\bar{U'}_{R}\gamma_{\mu}U'_{R} + \bar{U'}_{L}\gamma_{\mu}U'_{L}\right]B^{\mu},\nonumber \\  
\er 
\noindent while that for the $(2^a ger)$ we have 
\br 
 {\cal L}^{2^a\, ger}_{W^3,W^8} = -\frac{g}{2}&&\left[ \bar{U}_{2L}\gamma_{\mu}U_{2L} \left(W^{\mu}_{3}  - \frac{1}{\sqrt{3}}W^{\mu}_{8} \right)  +  \right. \nonumber \\  &&\left. \bar{D}_{2L}\gamma_{\mu}D_{2L}\left(-W^{\mu}_{3}  - \frac{1}{\sqrt{3}}W^{\mu}_{8} \right) + \right. \nonumber \\
 &&\left.  + \frac{2}{\sqrt{3}}\bar{D'}_{L}\gamma_{\mu}D'_{L}W^{\mu}_{8} \right] 
\er 
\br 
{\cal L}^{2^a\, ger}_{B} = \frac{g'}{2}&&\left[\bar{U}_{2L_{{}_{}}}\gamma_{\mu}U_{2L} + \bar{D}_{2L}\gamma_{\mu}D_{2L} + \right. \nonumber \\ &&\left. -\bar{U}_{2R}\gamma_{\mu}U_{2R}  + \bar{D}_{2R}\gamma_{\mu}D_{2R}  +  \right. \nonumber \\
&&\left.  + 3\bar{D'}_{R}\gamma_{\mu}D'_{R} + \bar{D'}_{L}\gamma_{\mu}D'_{L}\right]B^{\mu}. \nonumber \\  
\er 
\noindent In this case the couplings between the neutral  pions, $\Pi^{0}$ and $\pi^{0}_{1,2}$,  with $W^3, W^8$ and $B$ are
\br 
&&\!\! {\cal L}_{\pi^0_1 - W_3} = + i\frac{g}{2}f_{\pi_{1}}p_{\mu}W^{\mu}_{3}\,\,,\,\, {\cal L}_{\pi^0_2 - W_3} = + i\frac{g}{2}f_{\pi_{2}}p_{\mu}W^{\mu}_{3}\nonumber \\ 
&&\!\!  {\cal L}_{\pi^0_1  - B} =  -i\frac{g'}{2}f_{\pi_{1}}p_{\mu}B^{\mu}\,\,,\,\,{\cal L}_{\pi^0_2  - B} =  -i\frac{g'}{2}f_{\pi_{2}}p_{\mu}B^{\mu} \nonumber \\
&&\!\! {\cal L}_{\pi^0_1 - W_8} = - i\frac{g}{2}f_{\pi_{1}}p_{\mu}\frac{W^{\mu}_{8}}{\sqrt{3}}\,\,,\,\, {\cal L}_{\pi^0_2 - W_8} = - i\frac{g}{2}f_{\pi_{2}}p_{\mu}\frac{W^{\mu}_{8}}{\sqrt{3}}  \nonumber \\ 
&&\!\! {\cal L}_{\pi^0_1 - B}  = - i\frac{g'}{2}f_{\pi_{1}}p_{\mu}B^{\mu}\,\,,\,\,{\cal L}_{\pi^0_2 - B} = - i\frac{g'}{2}f_{\pi_{2}}p_{\mu}B^{\mu}\nonumber \\ 
&&\!\! {\cal L}_{\Pi - W_8}  = - i\frac{g}{2}(2F_{\Pi})p_{\mu}\frac{W^{\mu}_{8}}{\sqrt{3}}\,\,,\,\,{\cal L}_{\Pi - B} = - i\frac{g'}{2}(2F_{\Pi})p_{\mu}B^{\mu}. \nonumber \\ 
\label{neuc}
\er 
\begin{figure}[t]
\begin{center}
\epsfig{file=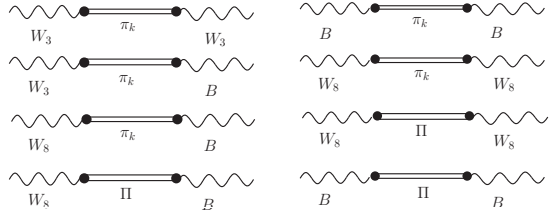,width=0.45\textwidth}
\caption{Contributions to the vacuum polarization $\Pi_{\alpha\beta}(p^2)$  of the  neutral gauge bosons. In this figure,  $k = 1, 2$ is the TC family index, and  for simplicity  we have not written the Lorentz indices.}
\end{center}
\end{figure}
\noindent In  Fig. II we represent the contributions of these couplings to the vacuum polarization tensor  of the  neutral gauge bosons.

 Therefore,  from the  contributions  depicted in  Fig. II,  and after considering  Eq.(\ref{neuc}), we can write the following mass matrix for neutral bosons in the base  $\{W_3, W_8 , B\}$
\br 
M^2_{neu} = \frac{g^2}{4} \left(\begin{array}{ccc} A_{W_3W_3} & 0   &  - A_{W_3B} \\  0  &  A_{W_8W_8}  &  A_{W_8B}   \\ -A_{BW_3}
  &  A_{BW_8}  &  A_{BB}  \end{array}\right)\nonumber  \\
\label{mass}
\er 
\noindent where   
\br 
&&A_{W_3W_3}= f^2_{\pi_{1}} + f^2_{\pi_{2}} \nonumber \\
&&A_{BW_3} = A_{W_3B} = t(f^2_{\pi_{1}} + f^2_{\pi_{2}})\nonumber \\
&&A_{W_8B} = A_{BW_8} = \frac{t}{\sqrt{3}}(f^2_{\pi_{1}} + f^2_{\pi_{2}}) + \frac{4t}{\sqrt{3}}F^2_{\Pi}  \nonumber  \\ 
&&A_{W_8W_8} = \frac{1}{3}(f^2_{\pi_{1}} + f^2_{\pi_{2}}) + \frac{4}{3}F^2_{\Pi}  \nonumber \\
&&A_{BB}=  2t^2(f^2_{\pi_{1}} + f^2_{\pi_{2}}) + 4t^2F^2_{\Pi}.  
\er 
\noindent and  we defined $t = \frac{g'}{g}$.  The eigenvalues of the matrix in  Eq.(\ref{mass}), assuming $F_{\Pi} >> f_{\pi_{1}},f_{\pi_{2}}$,  are then given by 
\br
&&M^2_{A} =0\,\,\,\,,\,\,\, M^2_{Z_{0}} \simeq\frac{g^2}{4}\left(f^2_{\pi_{1}} + f^2_{\pi_{2}} \right) \left[\frac{1 + 4t^2}{1 + 3t^2}\right] \nonumber \\ 
&&M^2_{Z'_{0}} \simeq \frac{g^2}{4}F^2_{\Pi}\left[\frac{4}{3} + 4t^2\right]. 
\er 
\noindent  The neutral physical states $(A_\mu, Z_{{0}_\mu}, Z'_{{0}_\mu})$ are the same described in  Ref.\cite{felice1} and $A_\mu $ represents the foton field. 

 In conclusion,   the gauge symmetry breaking in 3-3-1 models can be implemented dynamically because at the scale of a few TeVs  the $U(1)_X$
coupling constant becomes strong as we approach the peak existent in  Eq.(1)\cite{Das}\cite{doff}.  The exotic quark $T$  introduced in model will form a condensate  breaking  $SU(3)_{{}_{L}}\otimes U(1)_X$ to  electroweak symmetry.  In  this work we consider  only the T quark contribution to the $U(1)_X$ condensate, because we are assuming the most attractive channel(MAC) hypothesis\cite{Raby}. The MAC should satisfy  $\alpha_c(\mu_X)(X_{L}X_R) \sim 1$, and once $\alpha_c(\mu_X)$  is close to 1,  we can roughly estimate that $U(1)_X$ condensation
should occur only for the channel where $(X_{L}X_R) \gtrsim 1$. With the exception of the T quark, all 
other fermions have $(X_{L}X_R) < 1$. A more detailed analysis  requires a Schwinger-Dyson  equation calculation.
However, if contributions due to other channels, such as  those associated with  $U(1)_X$ condensation of  D or D', for example, we  expect  a  mass correction for the exotic  gauge bosons(or $F_{\Pi}$) not  larger than $20\%$, since $ \langle\bar{D}D\rangle \sim O(10^{-1})\langle\bar{T}T\rangle$\cite{Das}.

 In this brief report we explore the full  realization of the  dynamical symmetry breaking of the $SU(3)_{{}_{L}}\otimes U(1)_{{}_{X}}$ extension of the Standard Model\cite{tonasse} considering a model based on $SU(2)_{TC}\otimes SU(3)_{{}_{L}}\otimes U(1)_{{}_{X}}$. The electroweak symmetry is broken dynamically by a technifermion condensate  and we have determined the mass generated for the charged  gauge bosons of the model that result of the symmetry breaking.  

  We also  determine  the  mass matrix generated for the neutral gauge bosons of the model and found the same mass spectrum to the gauge bosons obtained with the introduction of fundamental scalars $\chi, \rho$ and $\eta$\cite{felice1}\cite{tonasse}. In other words, we verify  the equivalence between a  3-3-1 model  with a  scalar content formed by $\chi, \rho$ and $\eta$ , with a 3-3-1 model where the   dynamical symmetry breaking  is implanted by the system formed by the composite scalar bosons $\Phi^{}_{{}_{T}} = (\bar{T}t, \bar{T}b, \bar{T}T) \sim (\phi^{-}_{{}_{T}}, \phi^{--}_{{}_{T}}, \phi^{0}_{{}_{T}})$,  $\Phi_{{}_{TC(1)}} = (\bar{U}_1U_1 , \bar{U}_1D_1 , \bar{U}_1U') \sim (\phi^{0}_{{}_{TC(1)}}, \phi^{-}_{{}_{TC(1)}}, \phi^{+}_{{}_{TC(1)}}) $ and $ \Phi_{{}_{TC(2)}} = (\bar{D}_2U_2 , \bar{D}_2D_2 , \bar{D'}U_2) \sim (\phi^{+}_{{}_{TC(2)}}, \phi^{0}_{{}_{TC(2)}}, \phi^{++}_{{}_{TC(2)}})$. This system of composite bosons will produce the following hierarchical symmetry breaking $ SU(3)_{{}_{L}}\otimes U(1)_{{}_{X}}\stackrel{a}{\longrightarrow} SU(2)_{{}_{L}}\otimes U(1)_{{}_{Y}}\stackrel{b}{\longrightarrow}U_{em}$, with $a = \langle\phi^{0}_{{}_{T}}\rangle$ and $b = \langle \phi^{0}_{{}_{TC(1)}}, \phi^{0}_{{}_{TC(2)}} \rangle$. The novelty in this approach is that the minimal scalar content  is  fixed by the  condition  of the cancellation of  triangular  anomaly in TC sector. In Ref.\cite{dn} we discuss a mechanism for the dynamical mass generation in grand unified models with a horizontal symmetry,  incorporating quarks and techniquarks and including the generation of a large t quark mass. We expect that a  mechanism similar to the one  described in \cite{dn} can be developed for the  model  discussed here, this  possibility  and the determination of  the  mass spectrum of the composite Higgs bosons are topics that we intend to address in future work.


\begin{acknowledgments}
We thank A. A. Natale for useful discussions. This research was  partially supported by the Conselho Nacional de Desenvolvimento Cient\'{\i}fico e Tecnol\'ogico (CNPq).
\end{acknowledgments}



\begin {thebibliography}{99}
\bibitem{felice1}F. Pisano and  V. Pleitez, Phys. Rev. D{\bf46}, 410 (1992). 
\bibitem{frampton} P. H. Frampton, Phys. Rev. Lett. {\bf 69}, 2889 (1992).  
\bibitem{tonasse} V. Pleitez and M.D. Tonasse,  Phys. Rev. D{\bf 48}, 2353 (1993).
\bibitem{felice2} F. Pisano and  V. Pleitez, Phys. Rev. D{51 \bf}, 3865 (1995).
\bibitem{trecentes}  Alex G. Dias, C.A. de S.Pires and P.S. Rodrigues da Silva, Phys. Lett. {\bf B628}, 85 (2005);  Alex G. Dias,  C.A. de S.Pires,  V. Pleitez  and  P.S. Rodrigues da Silva, Phys. Lett. {\bf B621}, 151 (2005);  Alex G. Dias,  A. Doff, C. A. de S. Pires and P.S. Rodrigues da Silva,  Phys. Rev. {\bf D72}, 035006 (2005); Alex Gomes Dias, Phys. Rev. {\bf D71}, 015009 (2005); Alex G. Dias, J.C. Montero and  V. Pleitez, Phys. Lett. {\bf B637}, 85 (2006); Alex G. Dias and  V. Pleitez,   Phys. Rev. {\bf D73}, 017701 (2006), A. Doff, C. A. de S. Pires and  P. S. Rodrigues da Silva,  Phys. Rev.  {\bf D74},  015014 (2006).
\bibitem{dp1} A. Doff and  F. Pisano, Mod. Phys. Lett. {\bf A15},  1471 (2000).  
\bibitem{dp2} C.A de S. Pires and  O. P. Ravinez, Phys. Rev. {\bf D58},  035008 (1998); A. Doff and  F. Pisano, Mod. Phys. Lett. {\bf A14}, 1133 (1999); A. Doff and  F. Pisano, Phys. Rev.  {\bf D63}, 097903 (2001).
\bibitem{Das} Prasanta Das and Pankaj Jain, Phys. Rev. {\bf D 62}, 075001 (2000).
\bibitem{doff} A. Doff, Phys. Rev. {\bf D 76}, 037701 (2007).
\bibitem{lane} F. Sannino, hep-ph/0911.0931; K. Lane, {\it Technicolor 2000 }, Lectures at the LNF Spring School in Nuclear, Subnuclear and Astroparticle Physics, Frascati (Rome), Italy, May 15-20, 2000.
\bibitem{simmons} C. T. Hill and E. H. Simmons, Phys. Rept. {\bf 381}, 235 (2003) [Erratum-ibid. {\bf 390}, 553 (2004)]. 
\bibitem{Raby} S. Raby, S. Dimopoulos and L. Susskind, Nucl. Phys. {\b B169}, 373 (1980).
\bibitem{dn} A. Doff and A. A. Natale, Eur. Phys. J. {\bf C32}, 417 ,2003.


\end {thebibliography}
\end{document}